# Superior probabilistic computing using operationally stable probabilistic-bit constructed by manganite nanowire

Yadi Wang[1,2,†], Bin Chen[1,2,†], Wenping Gao[1,2], Biying Ye[1,2], Chang Niu[1,2], Wenbin Wang[1,3,4,5], Yinyan Zhu[1,4,5], Weichao Yu[1,4]*, Hangwen Guo[1,3,4,5]* and Jian Shen[1,2,3,4,5,6]*

[1] *State Key Laboratory of Surface Physics and Institute for Nanoelectronic Devices and Quantum Computing, Fudan University, Shanghai 200433, China*

[2] *Department of Physics, Fudan University, Shanghai 200433, China*

[3] *Shanghai Branch, Hefei National Laboratory, Shanghai 201315, China*

[4] *Shanghai Research Center for Quantum Sciences, Shanghai 201315, China*

[5] *Zhangjiang Fudan International Innovation Center, Fudan University, Shanghai 201210, China*

[6] *Collaborative Innovation Center of Advanced Microstructures, Nanjing 210093, China*

*Corresponding authors: wcyu@fudan.edu.cn; hangwenguo@fudan.edu.cn; shenj5494@fudan.edu.cn

†These authors contributed equally to this work.

# ABSTRACT

Probabilistic computing has emerged as a viable approach to treat optimization problems. To achieve superior computing performance, the key aspect during computation is massive sampling and tuning on the probability states of each probabilistic bit (p-bit), demanding its high stability under extensive operations. Here, we demonstrate a p-bit constructed by manganite nanowire that shows exceptionally high stability. The p-bit contains an electronic domain that fluctuates between metallic (low resistance) and insulating (high resistance) states near its transition temperature. The probability for the two states can be directly controlled by nano-ampere electrical current. Under extensive operations, the standard error of its probability values is less than 1.3%. Simulations show that our operationally stable p-bit plays the key role to achieve correct inference in Bayesian network by strongly suppressing the relative error, displaying the potential for superior computing performance. Our p-bit also serves as high quality random number generator without extra data-processing, beneficial for cryptographic applications.

**Keywords:** probabilistic computing, manganite nanowire, operational stability, electronic domain dynamics

# INTRODUCTION

Exploring the stochastic dynamics in nanomaterials has shown to be important in the emerging field of probabilistic computing, a new paradigm beyond conventional Von-Neumann architectures aiming to solve problems such as combinatorial optimization, invertible logic, and Bayesian inference [1-10]. The key ingredient of probabilistic computing is probabilistic bit (p-bit), a classical entity that can produce stochastic "0"s and "1"s with controllable probability. Unlike a CMOS-based p-bit that usually requires substantial number of transistors and is often not truly stochastic [11], nanomaterials with intrinsic fluctuation dynamics has emerged to offer great potential to construct a p-bit [1, 3, 4, 12-15]. Ideally, the nanomaterial involved should exhibit binary high (0) and low (1) resistance states that spontaneously fluctuate in time [16-20], and the fluctuation probability can be fine-tuned by external stimulus in a non-linear manner to allow efficient computation.

Various types of nanomaterials can be used to construct p-bits with tunable probability values whose stochastic features serve as an effective activation function (e.g. Sigmoid or Tanh function) [21, 22]. Yet in any experimental systems, instability of stochastic dynamics is always present so that broadening from the ideal activation function is inevitable, which can impede the performance on the computing tasks. While such broadening effect has been examined theoretically [23], a more important issue is the actual operational stability of a p-bit: whether the probability can return to the same value after experiencing extensive operations and how such stability impacts on the task performance. Such issue is critical since the input of each p-bit is determined by the collective outputs from other p-bits, all of which are fluctuating over time during computation [1, 24-26]. Owing to the sampling requirement, massive updating and tuning on its probability values takes place on individual p-bit, challenging its operational stability. Therefore, finding nanomaterials that can produce tunable probability with high operational stability is desirable to improve the performance of probabilistic computing.

In this work, we use an electronically phase separated (EPS) manganite nanowire to construct an operationally stable p-bit. In a simple two-probe geometry, we show direct control of two-state resistance fluctuations by nano-ampere input currents to achieve full-range probability tuning without assisting transistors. The probability curve fits well with the Sigmoid function – the basis of stochastic neurons in probabilistic computing. More importantly, we demonstrate that under extensive switching operations, the current-controlled p-bit shows exceptionally high stability with broadening of the probability values to be less than 1.3%. Numerical simulation results suggest that our operationally stable p-bit plays the key role to achieve correct inference in a Bayesian network by strongly suppressing the relative error as compared to p-bits with large standard errors, displaying the potential for superior computing performance. In addition, our nanowire is also capable of generating random numbers without any additional data processing, indicating the stable and true stochasticity with cryptographic quality.

## RESULTS

**Full electrically tunable probability in manganite nanowire**

We use manganite $(La_{2/3}Pr_{1/3})_{5/8}Ca_{3/8}MnO_3$ (LPCMO) thin film as a prototype material in this study to exploit the tunability on its fluctuation dynamics (see methods and supplementary Figure S1 for sample growth and characterization). The LPCMO system has been well known for its large-scale electronic domains whose dynamics is highly tunable by external stimuli [27-29]. When the system is spatially confined to the length scale of the EPS domain, the electronic transport is often governed by the dynamics of single residing domain which fluctuates between the ferromagnetic metallic (FMM) and charge ordered insulating (COI) states due to thermal instability near its phase transition point, as confirmed from experiments, theoretical model and simulations [30-32]. As shown in Figure 1a, LPCMO nanowire is ideal for such purpose, where a DC current source is applied as input, and the resistance values are monitored using a data acquisition system (see supplementary Figure S2 for nanofabrication details). Figure 1b shows the resistance vs temperature curve of the nanowire between 103 and 92 K where the overall nanowire is in the insulating background. When decreasing the temperature, clear resistance jumps on the order of 1 MΩ is observed with noticeable window of fluctuation towards lower resistance states. Such behavior can be attributed to a single domain that is dynamically fluctuating between the high resistance COI and the low resistance FMM states in a stochastic manner, which is sensitively detected by the transport measurements.

Next, we show that the stochastic domain dynamics can be well-controlled by directly tuning the input current using a simple two-probe geometry without any additional source of excitations. To demonstrate this behavior, we monitor the time-resolved resistance fluctuation at different input current levels. Figure 1d shows three representative examples. For an input current of 55 nA, the resistance of the nanowire spontaneously fluctuates between two resistance states but predominantly resides in the high-resistance state with small probability to transit into the low-resistance state. When increasing the input current to 61 nA, the domain fluctuates between the high and the low resistance states with nearly equal probability. When the input current is increased to 68.5 nA, the domain preferentially resides more in the low-resistance state. Here we define the probability $p$ as the fraction of the low-resistance state in the total collected data. Figure 1c plotted the probability $p$ vs input current curve. The probability can be continuously tuned from 0 to 1 when the input current is raised from 35 nA to 90 nA.

The physical origin of the current-tunable probability of the LPCMO nanowire can be attributed to either electric field effects or local Joule heating effects [33]. Here, we can exclude the local Joule heating effects by comparing the probability values at different temperatures and current (see supplementary Figure S3 and note). Therefore, our results point towards the electrical field driven local insulator-to-metal transition as the origin of the stochastic dynamics. Although the microscopic mechanism of the electric field driven phase transition is not fully understood, it is likely associated with current induced change of local charge distribution which can accurately control the energy difference between the COI and FMM states [34-37]. We note that the stochastic dynamics in our nanowire is fundamentally different from other Mott insulators such as $VO_2$ [38, 39]. In $VO_2$, each time a pumping voltage pulse is applied to trigger filamentary insulator-to-metal transition and induces stochastic relaxation [12]. In our nanowire, the electronic domain spontaneously fluctuates without any pumping voltages. The mechanism of ionic motion can also be excluded since the electric field to switch our stochastic domain is about $4\times10^2$ V/cm which is several orders of magnitude smaller than ionic-motion driven resistive switching in manganites [40]. We further examined the average rate of switching to be in between 3-4 ms which is consistent with previous reports, further confirming the intrinsic electronic domain dynamics as physical origin of our observation (Supplementary Figure S4). In any case, the key observation here is that the electric current can actively control a single domain's stochastic dynamics under dimensionality confinement conditions.

For probabilistic computing, a stochastic physical system that can represent sigmoid-type activation neuron is favored [21, 41]. We fit Figure 1d with sigmoid function $\text{Sigmoid}(x) = \frac{a}{1+e^{-k(x-x_c)}}$ (black curve), with the fitting parameter $a = 0.97529, k = 0.21856, x_c = 60.85965$, and the determination coefficient calculated to be 0.9996. An excellent match between the fitting curve and experimental results is observed, suggesting the high quality of our nanowire to serve as p-bit. Moreover, Figure 1c shows the probability when the input current is decreased from 90 nA to 35 nA (blue symbols), which coincides well with the probability acquired when current is increased from 35 to 90 nA (red symbols), further demonstrating the stability of domain fluctuation dynamics under cycles of input current scan.

In addition, the nature of strong correlation between multiple degrees of freedom in manganites allows us to control the p-bit by various external stimulus. We explored the effects of magnetic field and temperature on manipulating the p-bit behavior. Supplementary Figure S5 shows the $p$ values as function of external magnetic field and temperature. In both cases, full range tuning on the probability values between 0 and 1 is achieved. Such multi-field tunability allows us to implement electrical-current tunable probability at flexible temperatures and magnetic field (Supplementary Figure S6). We note that there are a few intermediate points between 0 and 1 as shown in Figure 1c which indicates partial domain switching. These points are statistically negligible that will not affect the probability determination of our p-bit (Supplementary Figure S7).

**P-bit with high operational stability**

Due to the presence of material impurity or the instability from intrinsic nonlinear dynamics in real physical system, broadening of the probability curve is inevitable [3, 4, 12, 42, 43], that is, a p-bit cannot maintain the same probability value over time under a fixed working condition. Our work here focuses on solving a more general issue: a p-bit no longer outputs with the same probability if its operational condition has been changed and switched back, e.g., $p(I_1) \neq p(I_1 \rightarrow I_2 \rightarrow I_1)$. This issue is critical since during the running phase of probabilistic computing, the working condition of each p-bit is determined by the collective outputs from other p-bits. Since all the p-bits are fluctuating over time, the input as well as the operational condition of an individual p-bit experiences massive updating. Such operational instability may cause vital damage to the performance of a probabilistic computer. Therefore, it is imperative to test the accuracy and stability of our p-bit under repeated switching of input signals. To quantify the operational stability of our LPCMO nanowire-based p-bit, we picked nine current values (52.5, 55, 57, 59, 61, 63, 65.5, 68.5, 73 nA) as switchable inputs according to the $p$ vs $I$ curve in Figure 1c which is again plotted in the x-y plane in Figure 2a. For each input, we collected the resistance vs time data for 60 seconds to calculate the probability $p$. Each input experienced 100 switching events and the $p$ values are counted (detailed procedure can be found in Method section). The resultant distribution of $p$ values is binned, and the histogram is depicted in Figure 2a. It can be seen that the broadening of the probability distribution is quite narrow against multiple switching events, indicating excellent stability of our p-bit under random operations. Figure 2b shows the detailed broadening distribution for three representative input currents (61, 65.5, 73 nA). Gaussian-type distributions are observed in all cases as shown in the fitting curves.

To quantify the broadening behavior, the full width at half maximum (FWHM) of the Gaussian peak was calculated to benchmark the accuracy of our nanowire-based p-bit. As shown in Figure 2c, the FWHM is largest for input current of 61 nA with the $p$ value around 0.5 and decreases continuously with the $p$ value approaching to 0 or 1. These results suggest high stability of our nanowire-based p-bit for probabilistic computing implementation under extensive sampling or cycling. Furthermore, our nanowire shows long endurance after being operated over extended amount of time. Supplementary Figure S8 shows the $p$ vs input current curve after 14 days over extensive current-switching operations. The resultant curve is nearly identical to the initial one in Figure 1c, illustrating the high endurance of our p-bit.

We simulate the probabilistic dynamics of a p-bit that exhibit operational instability according to the model [44] (details in Method section) that $m(t) = \text{sgn}[\text{rand}(-1,1) + \tanh[\beta I(t)] + \xi(t)]$ with discretized time step $t$, effective inverse temperature $\beta$ and dimensionless control parameter $I(t)$, e.g., electric current flowing through the p-bit. The operational variation of the p-bit is described by a random noise $\xi(t) \sim N\left(0, \frac{R}{Z^{|I(t)|}}\right)$ for $I(t) \neq I(t-1)$ and $\xi(t) = \xi(t-1)$ for $I(t) = I(t-1)$, which follows the normal distribution with a mean of zero and variance determined by broadening factors $R$ and $Z$. The fitting results are shown in Figure 2c, which yields $R = 0.0007$ and $Z = 7.9983$ agreeing well with experimental data. We also examined the behavior of two p-bits with $R = 0.05$ and $0.1$ as shown in Figure 2d. Compared to the operationally stable p-bit ($R = 0.0007$) with maximum standard error of 1.3%, clear broadening from expected value $\overline{m(t)}$ is observed. The maximal standard errors for p-bits of $R = 0.05$ and $0.1$ are 11% and 16%, respectively. Such level of standard error has been typically observed in several physical systems such as phase-changed [3, 12], RRAM [4] and stochastic MTJ-based p-bits [42, 43], as summarized in Supplementary Table 1. Therefore, our work provides the metric to compare the operational stability on multiple physical implementations of p-bits and its influence on the functionalities of Bayesian inferences as discussed below.

**Performance of operationally stable p-bit in Bayesian inference**

We then demonstrate via simulation that our nanowire-based p-bit shows superior performance in Bayesian inference, a prototype task in probabilistic computing [24, 45, 46]. Bayesian network is essentially a directed graphical model and is capable of performing tasks like Bayesian inference [21, 24]. Figure 3a shows the diagram of a genetic circuit with four generations. Without loss of generality, it is simply assumed that the children (such as $C_1$ and $C_2$) gets half their genes from their parents (such as $F_1$ and $M_1$). Hence, the correlation $g_{\text{Baye}}$ between $C_1$ and $C_2$ is 50% according to Bayesian theory (See note in Supplementary information). Similarly, we can calculate the correlation between more distant family members, such as Aunt $M_1$ and her niece $C_3$, resulting in 25% correlation.

We construct a p-circuit where each node is represented by a hardware p-bit, interconnected to model genetic influences. The correlation between two nodes, $C_1$ and $C_2$, both of which are bipolar variables with allowed values of -1 and +1, can be estimated statistically by:

$$g_{\text{stas}} = \frac{1}{N} \sum_{i=1}^{N} (C_1)_i (C_2)_i, \tag{1}$$

where $N$ represents the sample size in a single statistical measurement. During simulation, we make $M$ independent statistical measurements in total, with each measurement comprising $N = 10000$ samples of $C_1$ and $C_2$. The relative error $\delta$ is obtained by comparing the statistical results for all $M$ measurements and the one predicted by Bayesian theory, which is depicted in the Figure 3 and defined as

$$\delta = \frac{\left(\frac{1}{M} \sum_{i=1}^{M} (g_{\text{stas}})_i\right) - g_{\text{Baye}}}{g_{\text{Baye}}} \times 100\%. \tag{2}$$

Figures 3c and 3d show the $\delta$ for number of measurements $M = 1$ to 50. It is apparent that when $R = 0$ and $0.0007$, the fluctuations in $\delta$ are minimal, indicating that our p-bits can accurately estimate the theoretical values with the same number of measurements. However, when $R$ increases to 0.05 and 0.1, the fluctuations in $\delta$ significantly increase, leading to larger discrepancies between statistical results and theoretical values. All the data in Figure 3c and 3d converge with increasing number of measurements, resulting from the zero mean of random noise $\xi(t)$ as assumed in Eq. (2).

In Figure 3b, we further calculate the effective relative error, defined as $\delta_{\text{eff}} = \sqrt{\frac{\sum_{M=1}^{50} \delta_M^2}{50}}$ for all data demonstrated in Figure 3c and 3d. It is further evident that with increasing depth of the network, i.e. decreasing correlation between nodes, operationally instable p-bits could induce larger relative errors, indicating that the error can be amplified as the information propagates forward through a network with directed graph structure. Recent works show that the bidirectional information flow in undirected graph models (such as Restricted Boltzmann machines) may be immune from p-bits with broadening [23, 47]. However, it is still an open question on how a p-bit with operational instability play a role in an undirected graph model.

The example of Figure 3 demonstrates a simplest family diagram so that each node is affected by only two parent nodes. A more realistic task such as medical diagnosis usually requires multiple parent nodes, which requires a p-bit to work under multiple working conditions [48]. We show in the Supplementary information that the relative error $\delta$ will be further amplified if one increases the number of parent nodes (supplementary Figure S9 and S10). These results further elaborate the importance of the high stability of our experimental p-bit.

**High quality true random number generator**

When a p-bit is operated at probability of 0.5, it serves as a random number generator (RNG). While pseudo-RNG can be obtained via digital circuits, physical systems can acquire true-RNG since the randomness originated from intrinsic physical fluctuation is imperative for applications such as data encryption [49, 50]. A standard practice for evaluating the quality of a true RNG is carried out via statistical test suite (NIST STS, sp 800-22) [51]. Here we show that our p-bit can generate high-quality true random number without extra logic operations. Data with $p$ value of 0.5 in our manganite nanowire is used where we convert the resistance (voltage) values into bitstream of '0's and '1's without any further processing, and then feed to the NIST suite software. We examined 12 tests (which further subcategorized into 161 tests) in the suite and results are shown in Table 1. All the p-values are larger than 0.01 that passed the tests. Such behavior surpassed the quality of random number in most physical systems where extra data processing (such as XOR logic operations) are required [15, 17, 52, 53]. The high quality of the random number in our system can be directly attributed to the natural stability on the dynamical domain fluctuation which is potentially suitable for cryptographic applications. Here, 2 Hz sampling rate is used. Further exploration on how to increase the true randomness at smaller time scale will be an interesting topic [15].

## DISCUSSION

Finally, we discussed several aspects on how to implement our operationally stable p-bit for scalable probabilistic computing. Our nanowire shows a simple single layer geometry with size of 200 nm × 2 um which is suitable for building arrays of p-bits. Yet, its low working temperature presents a limitation for scalable devices with external electronic designs and circuits [2]. A possible solution and guideline is to raise the transition temperature using strain and doping effects. For example, in the manganites family, $La_{0.8}Ca_{0.2}MnO_3$ thin film exhibit strain-induced insulator-to-metal transition governed by sub-100 nm electronic domains at room temperature [54], $La_{0.67}Sr_{0.33}MnO_3$ thin films shows tunable insulator-to-metal transition temperature between 330 K to 400 K on certain substrates [55]. When nanofabricated, these systems can be promising candidates to explore room-temperature p-bits for scalable electronics. Another issue is the device-to-device variations which induces inevitable variability on the activation functions for different p-bits [1], even if each p-bit is operationally stable. Such effect may impair the performance of the computing tasks [23, 47]. A solution is to use extra circuits to compensate such variation [56]. In our case, the probability can be fully controlled by nano-ampere currents which may allow a power-saving solution on this matter.

In conclusion, we adopted a novel nanomaterial system to generate fully tunable probabilistic bit by controlling the electronic domain transition dynamics. The p-bit shows high stability during prolonged switching and operations which play critical role in the task of Bayesian inference. Also, our nanowire naturally serves as a high-quality random number generator without the assistance of additional logic operations, favorable for future cryptographic applications. Our results can also inspire future investigations on the importance of p-bit's operational quality on task performances.

## MATERIALS AND METHODS

**Thin film and nanowire fabrication**

40nm thick $(La_{2/3}Pr_{1/3})_{5/8}Ca_{3/8}MnO_3$ thin films were grown on $SrTiO_3$ substrates by pulsed laser deposition (248 nm, 1 J/cm$^2$ fluence) at 807 ℃ in $5 \times 10^{-3}$ Torr. The basic characterization results are shown in Supplementary Figure S1. E-beam lithography and Argon ion beam etching were applied to fabricate LPCMO nanowires with 200 nm line width. A pair of electrodes (Pd/Au) was patterned using E-beam lithography and evaporation with a 2 μm gap (Supplementary Figure S2).

**Electrical measurements**

DC electrical currents are supplied by Keithley 2450 source-meter. The time-resolved data is collected using National Instrument DAQ USB-6218 with sampling rate of 4.5 kHz. Measurement is taken in Quantum Design PPMS system. In random number tests, the sampling rate is 2 Hz and 200,000 data points are used. We note that remaining three tests require over one million data points which is not collected in the current experiments.

**Procedure for accuracy test**

For the results in Figure 2a, nine input current levels are selected (52.5, 55, 57, 59, 61, 63, 65.5, 68.5, 73 nA) which roughly corresponds to the probability from 0.1 to 0.9. The measurement procedure is listed as follow. (1) randomly choose an input current; (2) measure the resistance vs time data for 60 sec to calculate a $p$ value; (3) randomly switch to another input current and perform the same measurement in (2) to obtain another $p$ value; (4) repeat the process of (2) and (3) for 900 times so that each input current is counted for 100 times. The resultant $p$ value distribution is plotted in Figure 2a.

**Numerical simulation**

We develop a numerical model to quantitively describe the probabilistic dynamics of a p-bit that exhibit operational instability:

$$m(t) = \text{sgn}[\text{rand}(-1,1) + \tanh[\beta I(t)] + \xi(t)] \quad (3)$$

with bipolar output $m(t) \in \{-1,1\}$ at time step $t$. Here, $\overline{m(\beta I)}=\tanh(\beta I)$; and the probability for $m(t)$ to be at state 0 or 1 takes the form as $P(m(t) = 1|\beta I) = \text{sigmoid}(\beta I)$, effective inverse temperature $\beta$ and dimensionless control parameter $I(t)$, e.g., electric current flowing through the p-bit. The operational variation of the p-bit is described by a random noise $\xi(t)$ expressed as

$$\xi(t) \sim N\left(0, \frac{R}{Z^{|I(t)|}}\right), \quad I(t) \neq I(t-1) \quad (4)$$

that induces a random noise if the operational condition is changed, following the normal distribution with a mean of zero and variance determined by broadening factors $R$ and $Z$, wherein the denominator $Z$ modulates the dependence of variance on the input current $I$, and the numerator $R$ regulates the overall variance magnitude. Distinct from thermal noise, here the operational variance $\xi(t) = \xi(t-1)$ is assumed to be unchanged when the operation condition keeps the same between two steps, i.e., $I(t) = I(t-1)$. We perform 2000 statistical simulations using Eq. (3) and (4) with each simulation comprising 10000 sample points of $m$. For the simulation of Bayesian inference, $N$ =10000 samples were taken for each measurement where $C_1$ and $C_2$ are sampled simultaneously at each time step and $g_{\text{stas}}$ is calculated according to Eq. (1).

## SUPPLEMENTARY DATA

Supplementary data are available at NSR online.

## ACKNOWLEDGEMENTS

All authors are grateful for the fruitful discussions with Prof. Jiang Xiao, Prof. Zhongming Zeng and Prof. Zhe Yuan.

## FUNDING

This work was supported by the National Key Research Program of China (2022YFA1403300, 2020YFA0309100), the Innovation Program for Quantum Science and Technology (Grant No.2024ZD0300103), the National Natural Science Foundation of China (12074071, 12074073, 12204107), Shanghai Municipal Science and Technology Major Project (2019SHZDZX01), Shanghai Municipal Natural Science Foundation (22ZR1408100, 23ZR1407200) and Shanghai Science and Technology Committee (21JC1406200). Part of the experimental work was carried out in the Fudan Nanofabrication Laboratory.

## AUTHOR CONTRIBUTIONS

W.Y., H.G. and J.S. conceived and supervised the project. W.G., B.Y. prepared the samples. Y.W. and C.N. performed experimental measurements. B.C. performed the simulation and theoretical calculations. Y.W., B.C., W.Y., H.G. and J.S. wrote the manuscript with inputs and discussions from all the authors.

*Conflict of interest statement.* None declared.

## REFERENCES

1. Borders WA, Pervaiz AZ, Fukami S *et al.* Integer factorization using stochastic magnetic tunnel junctions. *Nature* 2019; **573**: 390-393.

2. Singh NS, Kobayashi K, Cao Q *et al*. CMOS plus stochastic nanomagnets enabling heterogeneous computers for probabilistic inference and learning. *Nature Communications* 2024; **15**: 2685.

3. Rhee H, Kim G, Song H *et al*. Probabilistic computing with NbOx metal-insulator transition-based self-oscillatory pbit. *Nature Communications* 2023; **14**: 7199.

4. Woo KS, Kim J, Han J *et al*. Probabilistic computing using $Cu_{0.1}Te_{0.9}$/$HfO_2$/Pt diffusive memristors. *Nature Communications* 2022; **13**: 5762.

5. Camsari KY, Faria R, Sutton BM, Datta S. Stochastic p-bits for invertible logic. *Physical Review X* 2017; **7**: 031014.

6. Kaiser J, Datta S. Probabilistic computing with p-bits. *Applied Physics Letters* 2021; **119**: 150503.

7. Si J, Yang S, Cen Y *et al*. Energy-efficient superparamagnetic Ising machine and its application to traveling salesman problems. *Nature Communications* 2024; **15**: 3457.

8. Misra S, Bland LC, Cardwell SG *et al*. Probabilistic Neural Computing with Stochastic Devices. *Advanced Materials* 2023; **35**: 2204569.

9. Li X, Wan C, Zhang R *et al*. Restricted Boltzmann Machines Implemented by Spin-Orbit Torque Magnetic Tunnel Junctions. *Nano Letters* 2024; **24**: 5420-5428.

10. Zhang Y, Zheng Q, Zhu X *et al*. Spintronic devices for neuromorphic computing. *Science China-Physics Mechanics & Astronomy* 2020; **63**: 277531.

11. Lin Y, Wang F, Liu B. Random number generators for large-scale parallel Monte Carlo simulations on FPGA. *Journal of Computational Physics* 2018; **360**: 93-103.

12. Valle Jd, Salev P, Gariglio S *et al*. Generation of tunable stochastic sequences using the insulator–metal transition. *Nano Letters* 2022; **22**: 1251-1256.

13. Wang K, Zhang YO, Bheemarasetty V *et al*. Single skyrmion true random number generator using local dynamics and interaction between skyrmions. *Nature Communications* 2022; **13**: 722.

14. Park TJ, Selcuk K, Zhang H-T *et al*. Efficient Probabilistic Computing with Stochastic Perovskite Nickelates. *Nano Letters* 2022; **22**: 8654-8661.

15. Vodenicarevic D, Locatelli N, Mizrahi A *et al*. Low-Energy Truly Random Number Generation with Superparamagnetic Tunnel Junctions for Unconventional Computing. *Physical Review Applied* 2017; **8**: 054045.

16. Hayakawa K, Kanai S, Funatsu T *et al*. Nanosecond random telegraph noise in-plane magnetic tunnel junctions. *Physical Review Letters* 2021; **126**: 117202.

17. Parks B, Bapna M, Igbokwe J *et al.* Superparamagnetic perpendicular magnetic tunnel junctions for true random number generators. *AIP Advances* 2018; **8**: 055903.

18. Gibeault S, Adeyeye TN, Pocher LA *et al.* Programmable electrical coupling between stochastic magnetic tunnel junctions. *Physical Review Applied* 2024; **21**: 034064.

19. Daniels MW, Madhavan A, Talatchian P *et al.* Energy-efficient stochastic computing with superparamagnetic tunnel junctions. *Physical Review Applied* 2020; **13**: 034016.

20. Schnitzspan L, Klaeui M, Jakob G. Nanosecond True-Random-Number Generation with Superparamagnetic Tunnel Junctions: Identification of Joule Heating and Spin-Transfer-Torque Effects. *Physical Review Applied* 2023; **20**: 024002.

21. Camsari KY, Sutton BM, Datta S. P-bits for probabilistic spin logic. *Applied Physics Reviews* 2019; **6**: 011305.

22. Cai JL, Fang B, Zhang LK *et al.* Voltage-Controlled Spintronic Stochastic Neuron Based on a Magnetic Tunnel Junction. *Physical Review Applied* 2019; **11**: 034015.

23. Zeng M, Li Z, Saw JW, Chen B. Effect of stochastic activation function on reconstruction performance of restricted Boltzmann machines with stochastic magnetic tunnel junctions. *Applied Physics Letters* 2024; **124**: 032404.

24. Faria R, Camsari KY, Datta S. Implementing Bayesian networks with embedded stochastic MRAM. *AIP Advances* 2018; **8**: 045101.

25. Debashis P, Ostwal V, Faria R *et al.* Hardware implementation of Bayesian network building blocks with stochastic spintronic devices. *Scientific Reports* 2020; **10**: 16002.

26. Shao Y, Duffee C, Raimondo E *et al.* Probabilistic computing with voltage-controlled dynamics in magnetic tunnel junctions. *Nanotechnology* 2023; **34**: 495203.

27. Zhang L, Israel C, Biswas A *et al.* Direct observation of percolation in a manganite thin film. *Science* 2002; **298**: 805-807.

28. Zhai H-Y, Ma J, Gillaspie DT *et al.* Giant discrete steps in metal-insulator transition in perovskite manganite wires. *Physical Review Letters* 2006; **97**: 167201.

29. Uehara M, Mori S, Chen C, Cheong S-W. Percolative phase separation underlies colossal magnetoresistance in mixed-valent manganites. *Nature* 1999; **399**: 560-563.

30. Ward TZ, Zhang X, Yin L *et al.* Time-resolved electronic phase transitions in manganites. *Physical Review Letters* 2009; **102**: 087201.

31. Ward TZ, Gai Z, Guo HW *et al.* Dynamics of a first-order electronic phase transition in manganites. *Physical Review B* 2011; **83**: 125125.

32. Dong S, Zhu H, Wu X, Liu JM. Microscopic simulation of the percolation of manganites.

*Applied Physics Letters* 2005; **86**: 022501.

33. Tokunaga M, Song H, Tokunaga Y, Tamegai T. Current oscillation and low-field colossal magnetoresistance effect in phase-separated manganites. *Physical Review Letters* 2005; **94**: 157203.

34. Guo H, Noh JH, Dong S *et al*. Electrophoretic-like gating used to control metal–insulator transitions in electronically phase separated manganite wires. *Nano Letters* 2013; **13**: 3749-3754.

35. Jeen H, Biswas A. Electric field driven dynamic percolation in electronically phase separated $(La_{0.4}Pr_{0.6})_{0.67}Ca_{0.33}MnO_3$ thin films. *Physical Review B* 2013; **88**: 024415.

36. Garbarino G, Acha C, Levy P *et al*. Evidence of a consolute critical point in the phase separation regime of $La_{5/8-y}Pr_yCa_{3/8}MnO_3$ (y~0.4) single crystals. *Physical Review B* 2006; **74**: 100401.

37. Dong S, Zhu C, Wang Y *et al*. Electric field induced collapse of the charge-ordered phase in manganites. *Journal of Physics-Condensed Matter* 2007; **19**: 266202.

38. del Valle J, Salev P, Tesler F *et al*. Subthreshold firing in Mott nanodevices. *Nature* 2019; **569**: 388-392.

39. Maher O, Jimenez M, Delacour C *et al*. A CMOS-compatible oscillation-based $VO_2$ Ising machine solver. *Nature Communications* 2024; **15**: 3334.

40. Sawa A. Resistive switching in transition metal oxides. *Materials Today* 2008; **11**: 28-36.

41. Zink BR, Lv Y, Wang J-P. Review of magnetic tunnel junctions for stochastic computing. *IEEE Journal on Exploratory Solid-State Computational Devices and Circuits* 2022; **8**: 173-184.

42. Funatsu T, Kanai S, Ieda Ji *et al*. Local bifurcation with spin-transfer torque in superparamagnetic tunnel junctions. *Nature Communications* 2022; **13**: 4079.

43. Schnitzspan L, Klaeui M, Jakob G. Electrical coupling of superparamagnetic tunnel junctions mediated by spin-transfer-torques. *Applied Physics Letters* 2023; **123**: 232403.

44. Chowdhury S, Grimaldi A, Aadit NA *et al*. A full-stack view of probabilistic computing with p-bits: devices, architectures and algorithms. *IEEE Journal on Exploratory Solid-State Computational Devices and Circuits* 2023; **9**: 1-11.

45. Shim Y, Chen S, Sengupta A, Roy K. Stochastic spin-orbit torque devices as elements for bayesian inference. *Scientific Reports* 2017; **7**: 14101.

46. Sebastian A, Pendurthi R, Kozhakhmetov A *et al*. Two-dimensional materials-based probabilistic synapses and reconfigurable neurons for measuring inference uncertainty using

Bayesian neural networks. *Nature Communications* 2022; **13**: 6139.

47. Morshed MG, Ganguly S, Ghosh AW. A deep dive into the computational fidelity of high variability low energy barrier magnet technology for accelerating optimization and bayesian problems. *IEEE Magnetics Letters* 2023; **14**: 6100405.

48. Harabi K-E, Hirtzlin T, Turck C *et al.* A memristor-based Bayesian machine. *Nature Electronics* 2023; **6**: 52-63.

49. Bikos A, Nastou PE, Petroudis G, Stamatiou YC. Random Number Generators: Principles and Applications. *Cryptography* 2023; **7**: 54.

50. Fukushima A, Seki T, Yakushiji K *et al.* Spin dice: A scalable truly random number generator based on spintronics. *Applied Physics Express* 2014; **7**: 083001.

51. Bassham LE, Rukhin AL, Soto J *et al.* A statistical test suite for random and pseudorandom number generators for cryptographic applications. 2010.

52. Ostwal V, Appenzeller J. Spin–orbit torque-controlled magnetic tunnel junction with low thermal stability for tunable random number generation. *IEEE Magnetics Letters* 2019; **10**: 1-5.

53. Song M, Duan W, Zhang S *et al.* Power and area efficient stochastic artificial neural networks using spin-orbit torque-based true random number generator. *Applied Physics Letters* 2021; **118**: 052401.

54. Kou Y, Miao T, Wang H *et al.* A strain-induced new phase diagram and unusually high Curie temperature in manganites. *Journal of Materials Chemistry C* 2017; **5**: 7813-7819.

55. Chromik Š, Štrbík V, Dobročka E *et al.* LSMO thin films with high metal–insulator transition temperature on buffered SOI substrates for uncooled microbolometers. *Applied Surface Science* 2014; **312**: 30-33.

56. Kaiser J, Borders WA, Camsari KY *et al.* Hardware-aware in situ learning based on stochastic magnetic tunnel junctions. *Physical Review Applied* 2022; **17**: 014016.

# Figures and Tables

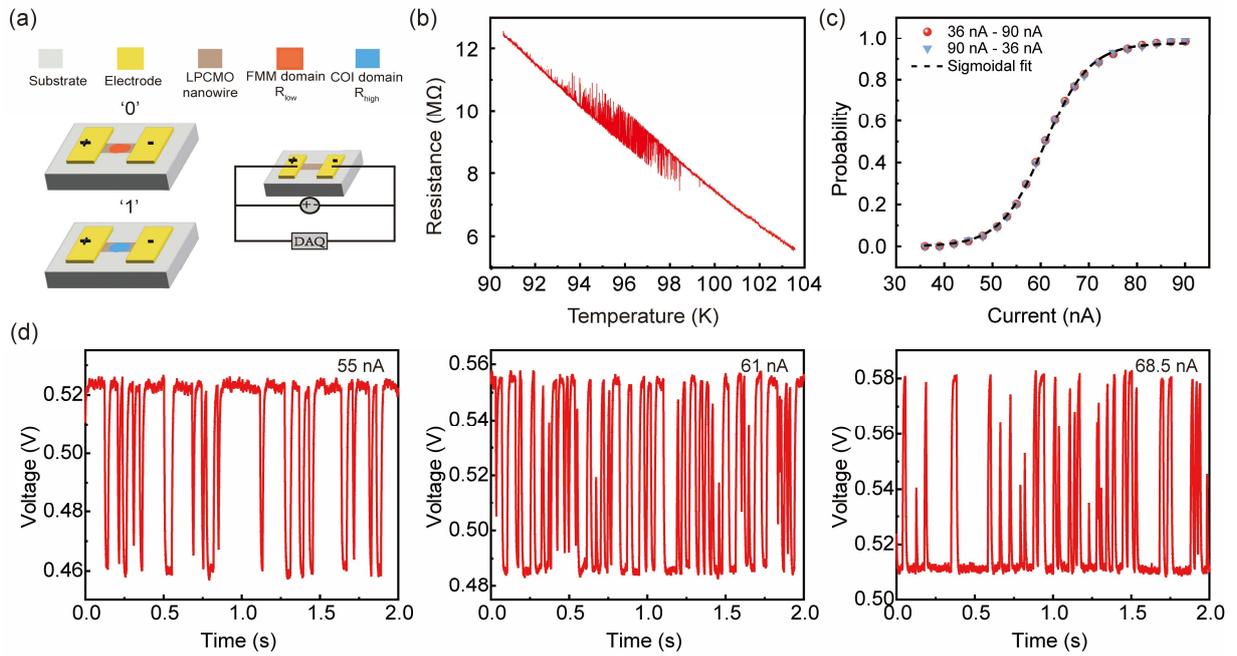

**Fig. 1. Fully tunable stochastic fluctuation behavior in manganites nanowire.** (a) schematic and measurement geometry of the LPCMO nanowire. The red and blue denotes a single domain that naturally fluctuates between FMM (red) and COI (blue) states under insulating background of the nanowire. (b) Temperature-dependent resistance curve showing stochastic single domain fluctuation in the nanowire. (c) full range probability (0 to 1) as function of input current measured at 97 K. Red and blue symbols represent up-scan and down-scan, respectively; the dashed line shows sigmoid fitting on the experimental results. (d) Resistance vs time curves under three distinct input current values.

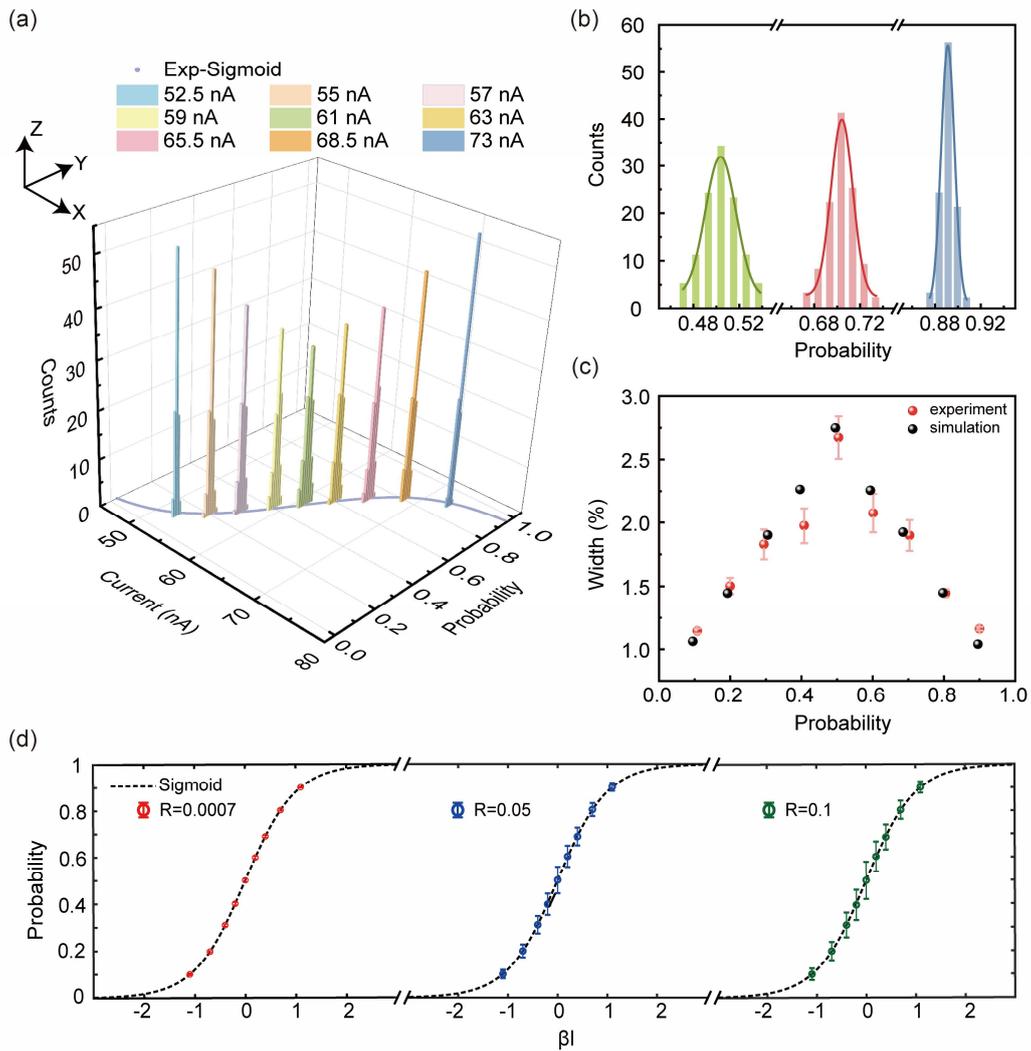

**Fig. 2. An operationally stable p-bit.** (a) Histograms depicting the probability distribution for nine individual input currents. *x-y* plane exhibits the current vs $p$ distribution as in Figure 1c; $z$ axis exhibits counts under multiple operations. For each count, a 60 sec data stream is recorded to calculate $p$. (b) Gaussian fitting of the probability distribution for input current of 61, 65.5, 73 nA. The solid line represents the Gaussian fit curve. (c) FWHM values for different input currents. (d) Probability curve simulated according to Eq. (1) with broadening parameter $R = 0.0007$ (LPCMO nanowire-based p-bit), $R = 0.05$ and $R = 0.1$. The circles in the figure represent the average value of the data, error bar represents the standard error.

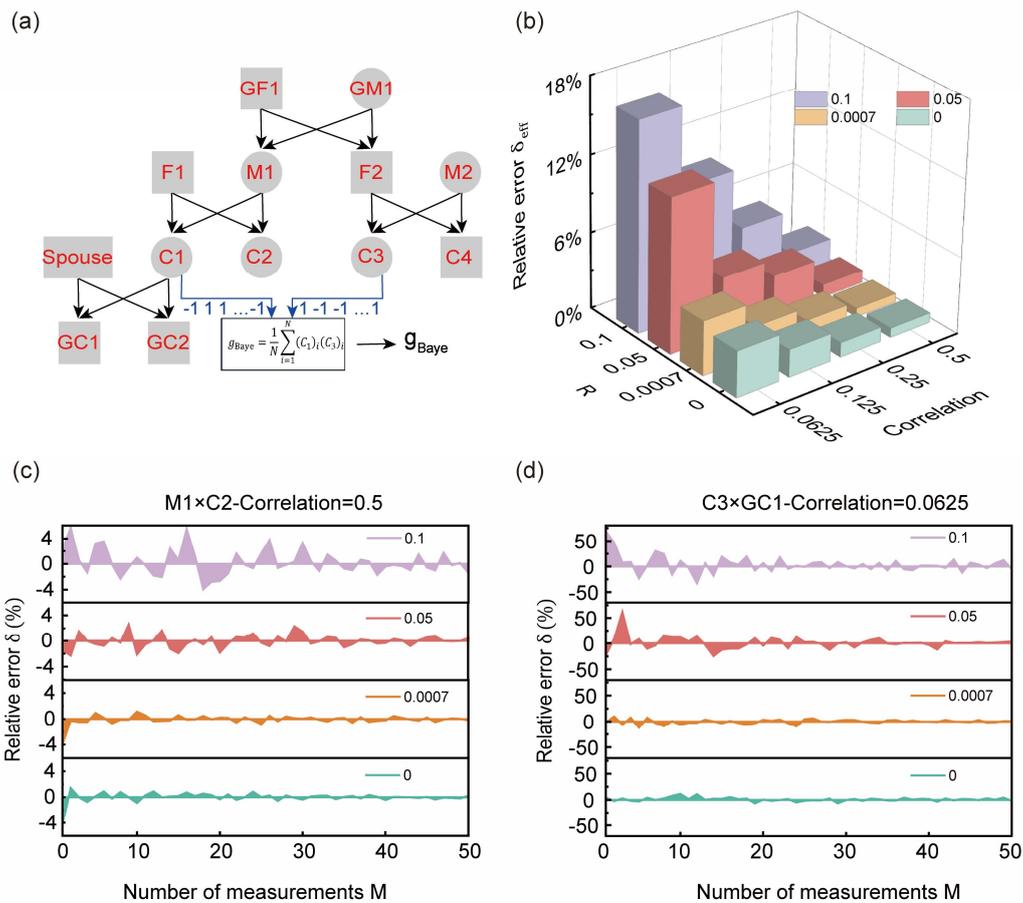

**Fig. 3. Performance of a p-circuit with operational instability.** (a) Schematics of a genetic diagram. (b) Performance of a p-circuit (effective relative error $\delta_{eff}$ between numerical statistics and Bayesian theory on correlation) vs depth of genetic circuits and broadening factor $R$. (c) & (d) Convergence of relative error $\delta$ for M1-C2 (c) and C3-GC1 (d) correlations with number of independent measurements $M$ sampled by p-bits with different $R$.

| Test | p-value | Result |
| --- | --- | --- |
| Frequency | 0.350485 | PASS |
| Block Frequency | 0.911413 | PASS |
| Cumulative Sums | 0.534146 / 0.213309 | PASS |
| Runs | 0.911413 | PASS |
| Longest Run | 0.911413 | PASS |
| Rank | 0.350485 | PASS |
| FFT | 0.739918 | PASS |
| Non-overlapping Template | >0.017912 | PASS |
| Overlapping Template | 0.534146 | PASS |
| Serial | 0.350485 / 0.350485 | PASS |
| Linear Complexity | 0.739918 | PASS |
| Approximate Entropy | 0.122325 | PASS |

**Table 1.** The results of the p values from NIST STS tests. The raw data passed 12 tests without XOR operation. Test passed if $p > 0.01$.